\documentclass[aps,prc,twocolumn,groupedaddress,showpacs,fleqn]{revtex4}

\usepackage{graphicx}
\usepackage{bm}
\usepackage{mathrsfs}
\begin{document}
\title{Luneburg-lens-like structural Pauli   attractive 
core   of nuclear force\\ at short distances 
 }

\author{S. Ohkubo }
\affiliation {Research Center for Nuclear Physics, Osaka University, 
Ibaraki, Osaka 567-0047, Japan }

\date{\today}

\begin{abstract}

The  nuclear force has been understood to have a repulsive  core at short distances, 
similar to a molecular force, since Jastrow proposed it in 1951.
 The existence of the repulsion was experimentally confirmed from the proton-proton
scattering $^1S_0$ phase shift, which becomes negative beyond 230 MeV.
This repulsion is essential for preventing the nucleon-nucleon system
 from collapsing by attraction.
The origin of the repulsion  has been considered to be due to the
 Pauli principle, similar to the repulsion originally revealed in $\alpha$-$\alpha$ scattering,
in many studies  including recent lattice QCD calculations.
  On the other hand, very recently it was shown that an inter-nuclear potential including 
 $\alpha$-$\alpha$ interactions  has a {\it Luneburg-lens-like  attraction} 
at short distances rather than  repulsion. We show that
 the nuclear force with an attractive potential at short distances that reproduces 
the experimental phase shifts  well
has a Luneburg-lens-like  {\it  structural Pauli attractive core}
 (SPAC)  at short distances and acts as  apparent {\it repulsion}. 
 The apparent  repulsion   is caused by  the deeply embedded unobservable Pauli forbidden 
  state  similar to  nucleus-nucleus potentials.

\end{abstract}

\maketitle

\par
In this paper it is shown 
the nuclear force with an attractive potential at short distances that reproduces
 the experimental phase shifts  well, has a Luneburg-lens-like structural 
{\it Pauli attractive  core} similar to the nucleus-nucleus potential and 
  acts as  apparent {\it repulsion}.
 This  study was inspired by   the recent discovery   of
 the Luneburg-lens-like  {\it structural   Pauli attraction} in  nucleus-nucleus
 potentials \cite{Ohkubo2016}.

\par
 The nuclear force is essential for the existence of  nuclei  \cite{Bohr1969}.
 It binds nucleons,  which allows the stable existence of atoms and matter, therefore  life. 
The origin of the nuclear force  was theoretically revealed by Yukawa \cite{Yukawa1935}.
 The nuclear force  was extensively studied by the Japanese nuclear force group
 \cite{Taketani1951,Taketani1956,Taketani1967,Hoshizaki968} based on  the three-stage
 theory  of Taketani \cite{Taketani1951,Taketani1956}. 
Jastrow proposed the existence of short range repulsion at short distances
 \cite{Jastrow1951}, which was   supported  by the negative $^1S_0$ phase shift
 observed by 310 MeV proton-proton scattering \cite{Stapp1959}. 
As shown in Fig.~1, a tremendous number of  studies \cite{Taketani1956,
Jastrow1951,Stapp1959,Hamada1962,Reid1968,Tamagaki1968,%
AV18,CD-Bonn2001,Reid93,ESC04,NNdata}
show  that the nuclear force has a repulsive core (hard or soft)  at short distances
in the innermost region III, and is   attractive  in the intermediate range  region II 
 and in  the outermost     one pion exchange  potential (OPEP) region I. 
 Phenomenological   potential models proposed in the 1960s
  include the Hamada-Johnston (HJ) potential with a hard core \cite{Hamada1962},  
Reid soft core potential  \cite{Reid1968} and  Tamagaki's  Gaussian
3 range soft (G3RS) core potential  \cite{Tamagaki1968}.
 The modern  high-precision  potentials 
 fitting   many $NN$ data \cite{NNdata}  include   Argonne V18 \cite{AV18}, 
CD-Bonn \cite{CD-Bonn2001}, Reid93 \cite{Reid93} and ESC04 \cite{ESC04}, in which 
a repulsive core is   introduced phenomenologically. 
 The  origin of the repulsive core has remained     a challenging subject.
It has been ascribed to  heavy  meson exchanges
 \cite{Nambu1957} and    the Pauli principle due to the substructure  of the nucleon
 \cite{Otsuki1964,Otsuki1965,Machida1965,Tamagaki1967}.

\begin{figure}[b]
\includegraphics[keepaspectratio,width=7.0cm] {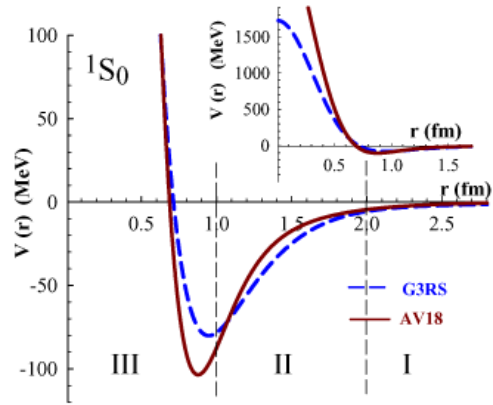}
 \protect\caption{\label{fig.1} {(Color online) 
Typical nuclear potentials for the $^1S_0$ channel, the   G3RS potential
  (dashed line)  \cite{Tamagaki1968} and the Argonne V18 potential (solid line)
  \cite{AV18}. 
}
}
\end{figure}

\par
After  QCD was established, new light was  shed on the
 origin  of the  repulsive core  from the quark model 
\cite{Neudatchin1977,Liberman1977,Obukhovsky1979,Faessler1982,Suzuki1983,Kukulin1984,Kukulin1985,Kukulin1992,Fujiwara2007}.
Neudatchin {\it et al.} \cite{Neudatchin1977} 
argued  that the repulsive core in the $S$ wave can arise from the  Pauli   forbidden
 state of the [42] orbital symmetry. 
Refs.\cite{Obukhovsky1979,Faessler1982} showed that the  color-magnetic quark-quark force  
 favors the  mixed symmetry state [42] acting attractively  and disfavors 
 the completely symmetric  orbital state [6] acting  repulsively. 
The two states can be almost degenerate \cite{Kukulin1992}, which means that 
 in  $S$ wave scattering the mixed  symmetry state can  contribute almost equally  as the 
 symmetric orbital state   in the inner region, III.
Ref.\cite{Suzuki1983} showed that the  repulsive  core of the  
 equivalent local potentials of   the resonating group method (RGM), which were derived 
using quark forces that cause  different  
admixtures of the mixed symmetry and WKB method,  largely originates from 
  the color-magnetic   exchange kernel. Recent lattice QCD calculations \cite{Inoue2010} 
reported that the repulsive core is due to the  Pauli principle 
\cite{Otsuki1964,Otsuki1965,Machida1965,Tamagaki1967,Neudatchin1977}.

\par 
The idea that  the repulsive core at short distances 
 comes from   the Pauli principle  \cite{Otsuki1964,Otsuki1965,Machida1965,Tamagaki1967}
 was originally inspired  by  analogy with the origin of the phenomenological
 repulsive core potential in  $\alpha$+$\alpha$ scattering. It was shown 
in Ref.\cite{Tamagaki1962} that  the {\it repulsive core} in  $\alpha$+$\alpha$ scattering, 
which is followed by an angular momentum ($L$)
 dependent {\it shallow  attraction} in the outer region, 
is a potential representation of the damped inner oscillations in the relative wave function
caused by  the Pauli principle  \cite{Tamagaki1965,Tamagaki1968E,Hiura1972}.  
 On the other hand,   it was also shown  later that not only  
$\alpha$+$\alpha$ scattering but also $\alpha$+$^{16}$O scattering
can be  well  reproduced by an $L$-independent  local {\it deep attractive} 
 potential {\it without} repulsive core  in which   the Pauli forbidden
 states of the RGM are embedded \cite{Kukulin1975,Buck1977,Ohkubo1977,Michel1983,Michel1998}.

\par
Very recently it has been  shown \cite{Ohkubo2016}  that the Pauli principle causes a 
Luneburg-lens-like structural {\it Pauli attraction}    in the 
internal region of the  nucleus-nucleus deep potential in contrast  to the traditional 
understanding that it causes a {\it repulsive core}
\cite{Tamagaki1962,Tamagaki1965,Tamagaki1968E,Hiura1972}. This was 
 demonstrated from the systematic study of nuclear rainbow scattering, prerainbows,
anomalous large angle scattering (ALAS), molecular structure and cluster structure 
\cite{Ohkubo2016}.  In a naive potential picture, the existence of   repulsion at short
 distances seems generally  indispensable  to prevent a system collapse by  attraction,
for example,  for  two atom molecules such as H-H.
 Historically, the observation of the  $S$ wave negative
 phase shifts  in   $\alpha$+$\alpha$ scattering \cite{Nilson1958} and  in proton-proton
 scattering \cite{Stapp1959} in the 1950s  lead  naturally  to   a repulsive core potential
at short distances based on  quantum  scattering theory of {\it structureless} particles.   
However, the recent finding of Ref.\cite{Ohkubo2016} urges us to ask whether a similar 
{\it Pauli attractive} core persists at short distances in a nucleon-nucleon potential
 given  that  the nucleon is composed of fermions.

\begin{figure}[t]
\includegraphics[keepaspectratio,width=7.6cm] {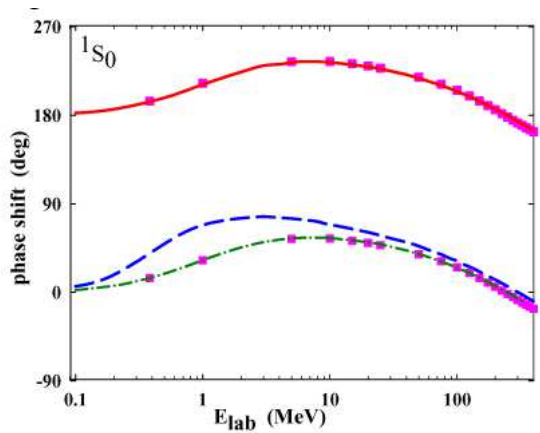}
 \protect\caption{\label{fig.2} {(Color online) 
 The  proton-proton scattering   $^1S_0$ phase shifts  calculated with
 the  SPAC potential  of Eq.~(2) (solid line),  the G3RS  potential   
 of Eq.~(1) (dashed line) and the  Reid93  potential \cite{Reid93} (dash-dotted line)
 are displayed.  The experimental data   (squares) are from  Ref.\cite{Arndt1983}.
 }
}
\end{figure}

\par
 From the   quark model viewpoint the Moscow  group 
\cite{Neudatchin1975,Kukulin1984,Kukulin1985,Kukulin1992}
 have been involved in developing  a model for such a nucleon-nucleon potential  
 that has an  attractive core due to the Pauli forbidden states.
They treated the region III and region II on the
same footing  phenomenologically representing it  either by a single Woods-Saxon potential, 
a single Gaussian potential or a single Yukawa potential.
 The apparent  core radius of the nuclear potential of Ref.\cite{Kukulin1984,Kukulin1985,Kukulin1992} 
is rather large extending to  near 1 fm \cite{Kukulin1992}. Also  underbinding  of 
 triton was pointed out \cite{Hahn1986}.
It is important to separate the region III and the established \cite{Fujiwara2007}
regions II and I.

\par
  We investigate  $^1S_0$  nucleon-nucleon  scattering where the complications due to
the spin and angular-momentum dependent forces such as  a tensor force  are absent.
The basic components of the modern  high-precision   potentials,
which have 40 (AV18) or similar number of  adjustable  parameters, are all present
 in the potential of HJ \cite{Hamada1962}, Reid \cite{Reid1968} and 
 G3RS \cite{Tamagaki1968}.
We take  the  G3RS potential (set $^1$E-1) \cite{Tamagaki1968}, which 
was modeled to reproduce the experimental phase shifts at $E_{lab}$=25-660 MeV 
by using a Gaussian function for the three regions
 as follows:
\begin{equation}
 V(r)  =  -5 e^{-(r/2.5)^2} -270 e^{-(r/0.942)^2}  +2000 e^{-(r/0.447)^2}.
\label{eq:G3RS}
\end{equation}
The strength of the potential is  in MeV and the range parameter is in fm.
The  phase shifts calculated using Eq.~(1) are displayed in  Fig.~2 by the  dashed line. 
 Our philosophy and prescription to find a {\it deep} potential is as
 follows. According to Ref.\cite{Ohkubo2016},  the $^1S_0$ phase shifts
would be equally well reproduced by replacing the {\it repulsive} core potential in Eq.~(1) by
 a {\it structural Pauli  attractive core} (SPAC), namely, by changing the sign of the strength 
 of the third core term (region III) of Eq.~(1).
The attractive first term (region I) and the second term (region II) of the SPAC potential 
in Eq.~(2)    correspond exactly to the first term (OPEP) and the second term 
(one-boson-exchange potential, OBEP) of Eq.~(1), 
respectively, which are based on the  established sound meson theoretical foundation
 \cite{Tamagaki1968}. The third term of the core (region III), repulsive in Eq.~(1) 
and attractive in Eq.~(2), is based on the theoretical foundation due to the Pauli principle.
\begin{equation}
 V(r)  =  -5 e^{-(r/2.5)^2} -270 e^{-(r/0.942)^2} -1850 e^{-(r/0.447)^2}.
\label{eq:SPAC}
\end{equation}
\noindent
It is surprising that a good  fit is easily obtained by a slight
 adjustment  to -1850 MeV.   
The phase shifts calculated  by the SPAC potential of Eq.~(2) 
are displayed in Fig.~2 by the solid line. Because of the generalized Levinson theorem, the phase shift
starts from 180$^\circ$ at $E_{lab}$=0 MeV.
The quality of fits to the experimental phase shifts is even better
 than the results  with the  G3RS potential, which cannot reproduce a  virtual  state  
near threshold  without reducing the height of the core. 
The SPAC potential is  almost phase shift equivalent to Eq.~(1).

\begin{figure}[t]
\includegraphics[keepaspectratio,width=7.0cm] {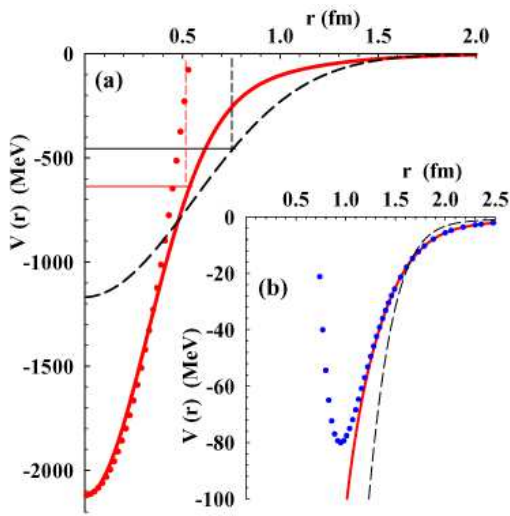}
 \protect\caption{\label{fig.3} {(Color online) 
 (a) The SPAC nuclear force with the Pauli attractive core at the short distances 
(red solid line) of Eq.~(2), Moscow potential (dashed line) of Ref.\cite{Kukulin1984}
 and the Luneburg-lens potential (red dotted line) are compared. 
The horizontal lines in panel (a) indicate the energy of the unobservable Pauli forbidden
  $0s$ state. The vertical dashed lines are to guide
 the eye. 
 (b) A magnified SPAC and Moscow potential in (a) is compared with the G3RS  potential 
(blue dotted line).
}
}
\end{figure}

\par
We investigate  whether   the  attractive core at short distances 
 is similar in  nature to  a Luneburg-lens-like potential.
A Luneburg lens \cite{Luneburg1964} 
is an aberration-free, spherically symmetric gradient-index lens, which
  decreases  radially from the  center to the outer surface $r=R$, and refracts
 all the parallel  incident  trajectories  to the focus $r=R_f$  ($< R$).
For such a lens the   refractive index $n$ is given by 
\begin{equation}
n^2(r \leq R) =  ({R_f^2-r^2+R^2})/{R_f^2}, \quad    n(r > R) = 1.
\end{equation}
The potential having this property \cite{Michel2002} is 
\begin{equation}
V(r \leq R) = V_0 \left(  {r^2}/{R^2}-1 \right),\quad V(r > R) = 0,
\end{equation}
\noindent where $V_0 = E (R/R_f)^2$ is the depth at $r=0$ with $E$ being
the energy of a material particle  moving in a potential $V(r)$.
\noindent This is   a harmonic oscillator (HO) potential truncated at $r=R$.
In Fig.3 the SPAC potential  is shown  in comparison 
the Luneburg-lens-like potential  together with the G3RS potential and Moscow potential
 of  Ref.\cite{Kukulin1984}.  The deeply bound unphysical Pauli forbidden $0s$ state 
is  indicated by the horizontal line in Fig.~3(a).
We see in Fig.~3(a) that the short distance  region of the nuclear potential 
 resembles the Luneburg lens with $V_0$=2120 MeV and $R$=0.54 fm. The attraction in
 the intermediate region II and the  outermost region I corresponds to the diffuse tail
 part of the potential, which  causes aberration \cite{Ohkubo2016}. 
The Luneburg-lens-like nature of the nuclear force  with the structural attractive
 core at short distances originates from the  third term of Eq.~(2).
The effect of the  potential of the  first and second  terms of
Eq.~(2) scarcely changes the Luneburg-lens-like origin of the core.
This can be understood analytically by the Taylor expansion of   Eq.~(2) 
to the first  order, which leads to  
  $V(r)= 2125((r/0.47)^2-1)$.
 The  2125 MeV and 0.47 fm are  close the  values of the above Luneburg lens parameters. 
The  third term of Eq.~(2) alone is well simulated  by a Luneburg-lens
 with $V_0$=1850 MeV and $R$=0.48 fm, which are close to the values
 $V_0$=1850 and $R$=0.447 fm derived from its   Taylor expansion. 
The Moscow potential is  considerably ``shallower'' than the SPAC potential
in the  core region, thus bringing a shallower Pauli forbidden state and a 
larger core radius.

\begin{figure}[t]
\includegraphics[keepaspectratio,width=8.6cm] {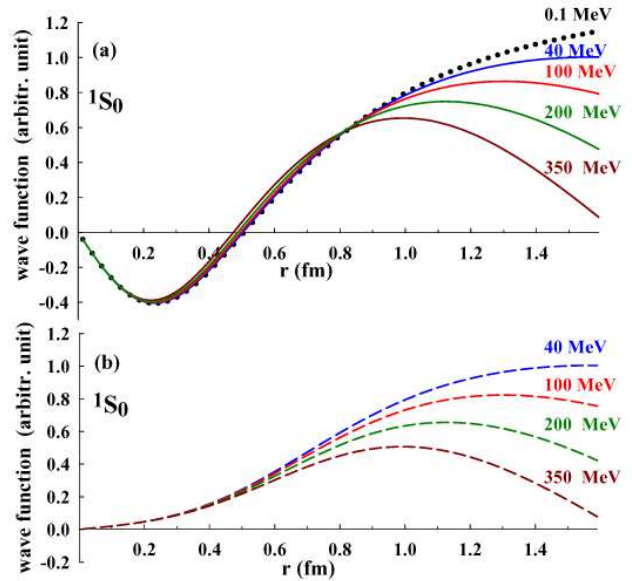}
 \protect\caption{\label{fig.5} {(Color online) 
 The calculated $^1S_0$ wave functions of proton-proton scattering at $E_{lab}$=0.1-350 MeV
 using (a)  the SPAC nuclear force potential with the 
Pauli attractive core at the short distances and (b) the G3RS potential 
 with the repulsive core at the short distances. The difference of wave functions in 
 (a) and (b) are seen in the core  region $r<$0.5.
}
}
\end{figure}

In Fig.~4   wave functions for proton-proton  scattering calculated using
 Eq.~(1) and Eq.~(2) are displayed.  One sees in Fig.~4(a) that the wave functions have
  a node at around $r=$0.5 fm for any incident energies. This shows that  the
  $s$ waves  are forced to be  orthogonal to the Pauli  forbidden $0s$ state deeply 
 embedded in the potential indicated in Fig.~5(a). One also notices that  the amplitudes
 of  the wave functions at the short distances  are  damped.
The node plays the role of preventing penetration of the wave functions into the region
$r < 0.5$ fm, namely,  collapsing of the system. This role is similar to the 
repulsive  core  at short distances in Fig.~1. 
Now it is clear that the Luneburg-lens-like structural Pauli attractive core plays the
 role of  apparent repulsion for any energy via the embedded Pauli forbidden state.
  From Fig.~4(a) and  Fig.~5(a), one sees that the   Luneburg-lens radius $R$
corresponds well to  the energy-independent nodal position $r\approx$ 0.5 fm. 
In Fig.~4(b) the wave functions for proton-proton scattering calculated  using the G3RS
 potential are displayed. As far as the asymptotic behavior  is concerned,
 the two wave functions calculated with the G3RS  repulsive core  potential 
 and the  SPAC  potential are ``phase shift  equivalent''.  However,
 while the wave functions are strongly damped at short distances for the repulsive core 
potential,  for the SPAC potential they survive with  non-vanishing significant amplitudes
of  the {\it inner oscillations}  at   short distances.  
One sees that the energy-independent nodal position in Fig.~4(a) corresponds well to the 
   the repulsive core radius at around $r=0.5$ fm  in Fig.~4(b).

\par 
In Fig.~5 the SPAC  nuclear force potential is compared with the  $\alpha$-$\alpha$ potential. 
Similar to the   nucleon composed of  three quarks, the  $\alpha$ 
particle is composed of  tightly bound four fermions and the interaction 
 is well described by a  deep potential between structureless point particles. 
In Fig.~5(d), the deep potential at short distances closely  resembles the Luneburg-lens truncated 
H.O. potential  indicated by the dotted lines. The overlap of the calculated 
deeply bound $0s$ and $1s$ states  with the H.O. wave functions  of the Pauli
 forbidden  states of the RGM, is 1 as was shown in Ref.\cite{Ohkubo2016}. 
Therefore the  deeply bound  states embedded in this potential  play the role of
 the Pauli forbidden states. The physical  $0^+$ state
 is forced to be orthogonal to them, by which the wave function has  two nodes
 as seen in Fig.~5(c). The outermost node at around $r=2$ fm, which arises  due to the 
orthogonality, corresponds  to the repulsive core radius of the   shallow $\alpha$-$\alpha$ 
potential.  The  situation of the $NN$ system is very similar to  $\alpha$-$\alpha$.
The  solved eigenfunction of the deeply  embedded bound  $0s$ state  at about $-637$ MeV,
which is indicated in  Fig.~5(b),  is also very similar  to the $0s$ wave function of the 
H.O. potential. This means that the three quarks are likely to be  confined in
 a harmonic oscillator potential. The deeply bound $0s$ state
 plays the role of the Pauli forbidden state of the RGM, similar to the $\alpha$-$\alpha$
 system.    In agreement with experiment, neither a physical bound state  nor a   resonant state 
appear in the  $^1S_0$ channel. In Fig.~5(a) the wave function
displayed is a virtual state obtained in the bound state approximation to show that the
 node appears   at around $r=0.5$ fm by the orthogonality to the $0s$ Pauli 
forbidden state. It was demonstrated mathematically in Ref.\cite{Ohkubo2016} that 
the Luneburg-lens-like attractive potential is a manifestation  of the  Pauli principle.

\begin{figure}[t]
\includegraphics[keepaspectratio,width=8.7cm] {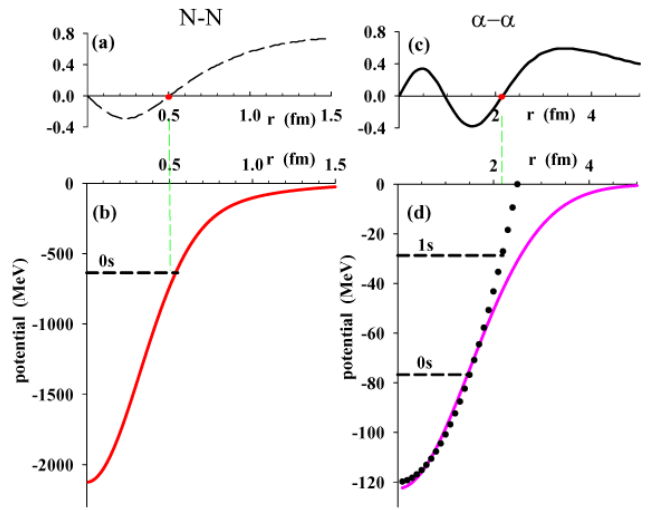}
 \protect\caption{\label{fig.4} {(Color online) 
The potentials and the wave functions (in arbitrary units) for the  nucleon-nucleon 
 and $\alpha$-$\alpha$ systems are compared. The SPAC $^1S_0$ $NN$  potential (b)
 with the structural Pauli attractive core at short distances and the  wave function (a)
are displayed in   comparison with  the $\alpha$-$\alpha$ deep potential with a structural
 Pauli  attraction of Ref.\cite{Ohkubo2016} (d)  and the Pauli allowed $s$ wave  function (c).
The horizontal dashed lines in panels (b) and (d) indicate the energy of the 
unobservable Pauli forbidden states 
  embedded in the $NN$ and $\alpha$-$\alpha$ potentials.} The  wave function
of $1s$  in (a)     and  $2s$   in (d) have been  calculated using the bound  state
 approximation. 
}
 \end{figure}

\par 
 It is  well known in nucleus-nucleus potentials that an $L$-independent  deep potential
 and an $L$-dependent  shallow potential with a repulsive core are interrelated. The latter
 is derived  phenomenologically \cite{Michel1985,Michel1988} or mathematically 
by supersymmetry theory \cite{Baye1987} from     the former but not vice versa. 
The widely used  Ali-Bodmer's $L$-dependent shallow $\alpha$-$\alpha$   potential with a 
repulsive core \cite{Ali1966} is  an approximate supersymmetry partner of the $L$-independent 
deep $\alpha$-$\alpha$ potential in which  the Pauli forbidden states are embedded
 \cite{Baye1987}.  Similarly the shallow nucleon-nucleon  potentials with a repulsive core 
 can be considered to be an approximate supersymmetric    partner 
of  the  deep SPAC   potential \cite{Michel1988}.
One is reminded that in the nucleus-nucleus potential case the differences between  the shallow
 and  deep potentials are clearly distinguished physically in the  observations 
  such as ALAS and nuclear rainbow, by which  shallow  potentials could not  survive. 
In the present case, the difference between the  wave functions at $r<$0.5  in Fig.~4(a) 
and (b)  may be seen in physical quantities  such as the binding energies in   
 few body systems. 
The underbinding problem for  triton  using a wide variety of  
modern $NN$ interaction models with a repulsive core  is well known 
\cite{Carlson1998}, which  has been ascribed to   three-body forces.
It is also to be noted that any   high-precision nuclear forces   with a repulsive core
cannot explain the existence of the recently  observed tetraneutron \cite{Kisamori2016}
   without  inconsistent modifications such as to introduce a remarkably 
attractive three-body \cite{Hiyama2016}.
 The non-vanishing  amplitudes of the inner oscillations
are expected to  give a significant energy gain for the 
 binding of   three and  four nucleon systems.

\par
  Although a shallow nucleus-nucleus potential prevailed    in the past decades 
 \cite{Hodgson1978},  it is now definitely agreed \cite{Horiuchi2012} that  a nuclear potential is deeply 
 attractive at short distances \cite{Horiuchi1991,Brandan1997,Michel1998,Khoa2007,Ohkubo2016}, 
 which is due to the Pauli principle \cite{Horiuchi1991,Ohkubo2016}.
On the other hand, the concept of baryon-baryon interaction with an attractive 
  deep potential  at short distances  
is   unfamiliar probably because  the fundamental nuclear model  and theory 
 were   developed using a  shallow potential with a repulsive core 
 \cite{Bohr1969,Brueckner1954}. 
   Ref.\cite{Inoue2010} reports that an deep attractive
 potential   appears in the {\bf \{1\}} representation $^1S_0$ channel   
 of SU(3) {\bf  8}$\times$ {\bf 8} of the flavor octet baryon  with spin 1/2. 
 Oka and Yazaki reported  that the $\Delta$-$\Delta$ potential is attractive 
at short  distances \cite{Liberman1977}.  
 As for the $\omega$ meson theory of  the core, a recent  holographic   model 
 using a  D4-D8 brane  configuration \cite{Hashimoto2009} reports that the core originates
 from extra spatial dimension and that the one-boson-exchange potential of an 
$\omega$ exchange captures  merely a part of the towers of massive  mesons.

\par
To summarize,  
 it was shown that 
the nuclear force with an attractive potential at short distances that reproduces
 the experimental $^1S_0$ phase shifts  well, has a Luneburg-lens-like  
{\it structural Pauli attractive core} (SPAC) similar to the nucleus-nucleus potential \cite{Ohkubo2016}.
The attractive core is as deep as -1850 MeV so that the embedded unobservable deeply bound 
 $0s$  state is closely  similar to  the  Pauli  forbidden state.
The  SPAC strongly prevents penetration of  the wave function   into the core region,
 thus  playing  the role of apparent  {\it repulsion}.
The energy-independent node at around  $r=0.5$ fm with  damped inner oscillations 
 in the wave function corresponds  to the core radius and the   Luneburg-lens radius $R$. 
The wave function can penetrate into the core region  significantly with the
  inner oscillation in contrast to   the repulsive core potential.
The nuclear forces with a repulsive core  can be considered  to be an approximate
   supersymmetric  shallow potential partner of the SPAC potential 
like the $\alpha$-$\alpha$ system.
 
 \par
The author thanks the Yukawa Institute for Theoretical Physics, Kyoto University for
 the hospitality extended  during a stay in February 2016 where part of this work 
 was done.


\begin{thebibliography}{aa}
\bibitem {Ohkubo2016}
S. Ohkubo, Phys. Rev. C {\bf 93}, 041303(R) (2016).
\bibitem {Bohr1969}
A. Bohr and B. R. Mottelson: {\it Nuclear Structure}, Vol. I,  
(W. A. Benjamin,  New York, 1969).
\bibitem {Yukawa1935}
H. Yukawa, Proc. Phys.-Math. Soc. Japan. {\bf 17}, 48 (1935).
\bibitem {Taketani1951} 
M. Taketani, S. Nakamura, and M. Sasaki, Prog. Theor. Phys. {\bf 6}, 581 (1951).
\bibitem {Taketani1956}
M. Taketani,  {\it Meson Theory III, Nuclear Forces}, 
 Prog. Theor. Phys. Suppl. {\bf 3}, 1 (1956).
\bibitem {Taketani1967}
M. Taketani,  {\it Nuclear Forces I, Nuclear Forces in Dynamical Region},  
Prog. Theor. Phys. Suppl. {\bf 39}, 1 (1967).
\bibitem {Hoshizaki968}
N. Hoshizaki,  {\it Nuclear Forces II, Nuclear Forces in Core Region}, 
Prog. Theor. Phys. Suppl. {\bf  42}, 1 (1968). 
\bibitem {Jastrow1951}
 R. Jastrow, Phys. Rev. {\bf 81}, 165 (1951).
\bibitem {Stapp1959} 
H. P. Stapp, T. J. Ypsilantis, and N. Metropolis,
 Phys. Rev. {\bf 105}, 302 (1957).
\bibitem {Hamada1962} 
T. Hamada and  I. D. Johnston, Nucl. Phys. {\bf 34}, 382 (1962).
\bibitem {Reid1968} 
R. V. Reid, Ann. Phys.  {\bf 50}, 411 (1968).
\bibitem {Tamagaki1968} 
R. Tamagaki, Prog. Theor. Phys. {\bf 39}, 91 (1968).
\bibitem {AV18} 
R. B. Wiringa, V. G. J. Stoks, and R. Schiavilla,
Phys. Rev. C {\bf 51}, 38 (1995).
\bibitem {CD-Bonn2001} 
R. Machleidt, Phys. Rev. C {\bf 63}, 024001 (2001). 
\bibitem {Reid93} 
V. G. J. Stoks, R. A. M. Klomp, C. P. F. Terheggen, and J. J. de Swart,
Phys. Rev. C {\bf 49}, 2950 (1994).
\bibitem {ESC04} 
Th. A. Rijken, Phys. Rev. C {\bf 73},  044007 (2006).
\bibitem {NNdata}
http://nn-online.org/
\bibitem {Nambu1957}
Y. Nambu, Phys. Rev. {\bf 106}, 1366 (1957);
K. Holinde, Phys. Rep. {\bf 68},121 (1981).
\bibitem {Otsuki1964}
S. Otsuki, R. Tamagaki, and M. Wada, Prog. Theor. Phys. {\bf 32}, 220 (1964).
 \bibitem {Otsuki1965}
  S. Otsuki, R. Tamagaki, and M. Yasuno,
in {\it Commemoration Issue for the 30th of the Meson Theory by Dr. H. Yukawa},
  Prog. Theor. Phys. Suppl. {\bf  E65}, 578  (1965).
\bibitem {Machida1965}
S. Machida and M. Namiki, Prog. Theor. Phys. {\bf 33}, 125 (1965).
\bibitem {Tamagaki1967} 
R. Tamagaki, Rev, Mod. Phys. {\bf 39}, 629 (1967).
\bibitem {Neudatchin1977}
V. G. Neudatchin, Yu. F. Smirnov, and R. Tamagaki, Prog. Theor. Phys. {\bf 58}, 1072 (1977).
\bibitem {Liberman1977}
D. A. Liberman, 
 Phys. Rev. D {\bf  16}, 1542 (1977);
C. S. Warke and R. S. Shanker,
 Phys. Rev. C {\bf  21}, 2643 (1980);
J. E. F. T. Ribeiro, Z.Phys. C {\bf  5}, 27 (1980);
M. Oka and K. Yazaki, Phys. Lett. {\bf B 90}, 41 (1980);
M. Oka and K. Yazaki, Prog. Theor. Phys. {\bf 58}, 572 (1981);
I. T. Obukhovsky, A. M. Kusainov, Phys. Lett. {\bf B 238}, 142 (1990);
A. M. Kusainov, V. G. Neudatchin, and I. T. Obukhovsky,
Phys. Rev. C {\bf 44}, 2343  (1991);
M. Oka, K. Shimizu, and K. Yazaki, Prog. Theor. Phys. Suppl. {\bf 137}, 1 (2000).
\bibitem {Obukhovsky1979} 
I. T. Obukhovsky, V. G. Neudatchin, Yu. F. Smirnov, Yu. M. Tchuvil'sky,
Phys. Lett. {\bf B 88}, 231 (1979).
\bibitem {Faessler1982}
A. Faessler, F. Fernandez, G. L\"{u}beck, and K. Shimizu, 
Phys. Lett. {\bf B 112}, 201 (1982).
\bibitem {Suzuki1983} 
Y. Suzuki and K. T. Hecht, Phys. Rev. C {\bf 27}, 299 (1983);
Phys. Rev. C {\bf 28}, 1458 (1983);
Y. Suzuki, 
Nucl. Phys.  {\bf 430}, 539 (1984).
\bibitem {Kukulin1984} 
V. I. Kukulin, V. N. Pomerantsev, V. M. Krasnopol'sky, and P.B. Sazonov,
Phys. Lett. {\bf B135}, 20 (1984).
\bibitem {Kukulin1985} 
V.M. Krasnopol'sky, V.I. Kukulin, V.N. Pomerantsev, P.B. Sazonov,
Phys. Lett. {\bf B 165}, 7 (1985).
\bibitem {Kukulin1992}
V. I. Kukulin and V. N. Pomerantsev.
Prog. Theor.  Phys. {\bf 88}, 159 (1992) and references therein.

\bibitem {Fujiwara2007} 
Y. Fujiwara, Y. Suzuki, and C. Nakamoto, 
Prog. Part. Nucl. Phys. {\bf 58}, 439 (2007) and references therein.

\bibitem {Inoue2010}
T. Inoue, N. Ishii, S. Aoki, T. Doi, T. Hatsuda, Y. Ikeda,  K. Murano, and H. Nemura,
 Prog. Theor. Phys. {\bf 124}, 591 (2010).
\bibitem {Tamagaki1962} 
I.  Shimodaya, R. Tamagaki, and H. Tanaka,  
Prog. Theor.  Phys.    {\bf 27}, 793 (1962);
 {\it ibid.} {\bf 25}, 853 (1961).
\bibitem {Tamagaki1965}
 R. Tamagaki and H. Tanaka,  
Prog. Theor.  Phys.    {\bf 34}, 191 (1965).
\bibitem {Hiura1972} 
J. Hiura and R. Tamagaki,  Suppl.  Prog. Theor.  Phys. {\bf 52}, 25 (1972).
\bibitem {Tamagaki1968E} 
R. Tamagaki,  
Prog. Theor.  Phys. Suppl.   {\bf E68}, 242 (1968).
\bibitem {Kukulin1975} 
V. I. Kukulin, V. G. Neudatchin, and Yu. F. Smirnov,
Nucl. Phys.   {\bf A245},  429 (1975).
\bibitem {Buck1977} 
B. Buck, H. Friedrich, and C. Wheatley,
Nucl. Phys.   {\bf A275},  429 (1975).
\bibitem {Ohkubo1977} 
S. Ohkubo, Y. Kondo, and S. Nagata,   Prog. Theor. Phys.  {\bf 57},  82 (1977).
\bibitem {Michel1983}
 F. Michel,
 J. Albi\'{n}ski, P. Belery, Th. Delbar, Gh. Gr\'{e}goire, B. Tasiaux, and G. Reidemeister,
 Phys. Rev. C {\bf 28}, 1904 (1983).
\bibitem {Michel1998} 
F. Michel, S. Ohkubo, and G. Reidemeister, 
Prog. Theor. Phys. Suppl. {\bf 132}, 7 (1998) and references therein.
\bibitem {Nilson1958} 
R. Nilson, W. K. Jentschke, G. R. Briggs, R. O. Kerman, and J. N. Snyder,
Phys. Rev. {\bf 109}, 850 (1958).
\bibitem {Neudatchin1975} 
V. G. Neudatchin, I. T. Obukhovsky, V. I. Kukulin, and N. F. Golovanova,
Phys. Rev. C {\bf 11}, 128 (1975);
V. G. Neudatchin, N. P. Yudin, Yu. L. Dorodnykh, and I. T. Obukhovsky,
Phys. Rev. C {\bf 43},  2499  (1991).
\bibitem {Hahn1986}
K. Hahn, P. Doleschall, and E.W. Schmid,
Phys. Lett. {\bf B169} , 118 (1986);
S. Nakaichi-Maeda, Phys. Rev. C {\bf 34},  303  (1986).
\bibitem {Arndt1983}
R. A. Arndt, L. D. Roper, R. A. Bryan, R. B. Clark, B. J. VerWest, and P. Signell,
Phys. Rev. D {\bf  28}, 97 (1983).
\bibitem {Luneburg1964} 
  R. K. Luneburg, {\it Mathematical Theory of Optics} (University of California Press, 
California, 1964).
\bibitem {Michel2002}
F. Michel, G. Reidemeister, and  S.   Ohkubo, 
Phys. Rev. Lett. {\bf 89}, 152701 (2002).
\bibitem {Michel1985}
 F. Michel and G. Reidemeister,  J. Phys.  {\bf G 11}, 835 (1985).
\bibitem {Michel1988}
F. Michel, {\it Proceedings of the International Symposium on Developments
 of Nuclear Cluster Dynamics},(World Scientific, Singapore, 1989),
pp.148;
 F. Michel and G. Reidemeister, Zeit. Phys. {\bf 329}, 385 (1988).
\bibitem {Baye1987}
D. Baye,  Phys. Rev. Lett. {\bf 58}, 2738 (1987).
\bibitem {Ali1966} 
S. Ali and A. R. Bodmer, Nucl. Phys. {\bf A80}, 99 (1966).
\bibitem {Carlson1998}
J. Carlson and R. Schiavilla,
Rev.  Mod.  Phys. {\bf 70}, 743 (1998).
\bibitem {Kisamori2016}
K. Kisamori {\it et al.},
Phys. Rev. Lett. {\bf 116}, 044004 (2016).
\bibitem {Hiyama2016}
E. Hiyama, R. Lazauskas, J. Carbonell, and M. Kamimura,
Phys. Rev. C {\bf 93}, 044004 (2016).

\bibitem {Hodgson1978}
P. E. Hodgson, {\it Nuclear Reactions  and Nuclear Structure} (Clarendon Press, Oxford 1971);
P. E. Hodgson, {\it Nuclear Heavy Ion Reactions} ( Clarendon Press, Oxford, 1978).
\bibitem {Horiuchi2012} 
H. Horiuchi, K. Ikeda, and K. Kat\={o},
Prog. Theor. Phys. Suppl. {\bf 192}, 1 (2012).
\bibitem {Horiuchi1991} 
H. Horiuchi, {\it Trends in Theoretical Physics} Vol. 2,
 ed. P. J. Ellis and Y. C. Tang (Addison-Wesley, Redwood City, 1991), p. 277.
\bibitem {Brandan1997}
M.-E. Brandan and G. R. Satchler, Phys. Rep. {\bf 285}, 143 (1997).
\bibitem {Khoa2007} 
D. T. Khoa,  W. von Oertzen, H. G. Bohlen, and S.  Ohkubo, 
J. Phys. {\bf G 34}, R111 (2007).
\bibitem {Brueckner1954}
K. A. Brueckner, C. A. Levinsin, and H. M. Marmound,
Phys. Rev.  {\bf 95}, 217 (1954);
H. A. Bethe Phys. Rev. {\bf 103}, 1353 (1956);
J. Goldstone, Proc. R. Soc.  A {\bf 239}, 267 (1957).

\bibitem {Hashimoto2009}
K. Hashimoto, T. Sakai, and S. Sugimoto,
Prog. Theor. Phys. {\bf  122}, 427 (2009).

\end{thebibliography}
\end{document}